## In Memoriam J. Robert "Bob" Dorfman (1937–2025)

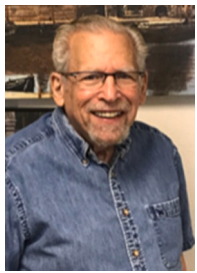

Jacob Robert Dorfman, "Bob", born May 20, 1937, died on August 27, 2025. He is survived by his children Theodore Dorfman, Ann Jamieson, and Sydney Passin, two stepchildren Lee and Ruth Futrovsky as well as numerous grandchildren. His beloved wife of many years, Celia Shapiro, passed away in early 2024. Bob was an internationally known physicist who made major contributions to statistical physics and especially kinetic theory over almost six decades.

Bob was a Ph.D. student of Ted Berlin at the Johns Hopkins University. His thesis in 1961 was entitled 'The Theory of Linear Response of Systems Subjected to External Forces'. Bob then followed Berlin to the Rockefeller University. Shortly thereafter E.G.D. "Eddie" Cohen moved to Rockefeller as Berlin's successor, and Bob and he started a long-lasting collaboration that led to numerous important results and papers.

At the time many groups were trying to understand how to extend the Boltzmann equation to higher order in the density. The earliest of these attempts was by Bogolyubov who assumed that an analytic expansion in the density was possible. Bob and Eddie showed that such an expansion was not possible due to quasi-long-ranged dynamical correlations that do not show up in classical equilibrium statistical mechanics. Among other things this work implied that transport coefficients of fluids do not have an analytic density expansion, which was a big surprise at that time.

Bob subsequently moved to the University of Maryland at College Park. He continued his work with Eddie. The next important contribution by them was a kinetic theory explanation of the unexpected long-time tails (LTTs) observed by Alder and Wainwright in molecular dynamics simulations of the velocity autocorrelation function, which determines the self-diffusion coefficient in fluids. Bob and Eddie showed that these LTTs are closely related to the divergence of the density expansion of the transport coefficients mentioned above, although the spatial correlations that cause them are hydrodynamic in nature and of even longer range.

While Bob and Eddie explained the LTTs using kinetic theory, a phenomenological mode-coupling theory giving the same results, but applicable at arbitrary densities, was developed by Matthieu Ernst, Eivind Hauge, and Hans van Leeuwen. Motivated by this connection Bob and Matthieu Ernst spent several fruitful years showing that for a variety of problems kinetic theory and mode-coupling theory always give identical results, when the mode-coupling theory is used in the limit where kinetic theory is valid. All this work showed that long-ranged spatial correlations cause the coefficients in local generalizations of Navier-Stokes hydrodynamic equations to not exist. In particular, due to LTT physics, generalized linear hydrodynamics is nonlocal in space and time, whereas nonlinear hydrodynamics contains non-analytic powers of the macroscopic gradients. These results were completely unexpected.

As interesting as all the work associated with LTTs was, the predicted effects were generally quite small and difficult to observe in real experiments. This all changed when Bob along with Ted Kirkpatrick and Eddie Cohen considered spatial correlations in non-equilibrium steady states (NESS) and, specifically, in a constant temperature gradient in a fluid. They predicted that light scattering in such a fluid would exhibit a very large increase at small wavenumber. This in turn reflects the fact that there are extraordinarily long-ranged spatial correlations in a fluid with a temperature gradient. This surprising result was first experimentally confirmed in detail by Jan Sengers and collaborators. The same phenomenon has turned out to be even more dramatic in liquid mixtures in the presence of a concentration gradient, and it is currently referred to as "giant nonequilibrium fluctuations".

Bob and Ted continued their collaboration by constructing kinetic equations for a dilute condensed Bose gas. Among other things, they were able to derive Landau's two-fluid hydrodynamic equations for a superfluid as well as the complicated thermodynamics that is associated with these equations. All this work was well before the experimental advances in dilute condensed Bose systems. The results of Bob and Ted were later re-derived by other workers in this field.

Bob was not only an exceptional physicist but also contributed to the University of Maryland by serving as Director of the Institute for Physical Science and Technology, then Dean of College of Mathematics and Physical Sciences, and eventually Provost of the University. This equally successful career in management was interrupted by a serious illness, which he managed to overcome successfully.

After this interlude, Bob came back to physics. Motivated by his long-standing interest in fundamental aspects of statistical mechanics, and inspired by the ideas of Pierre Gaspard and Thomas Gilbert, Bob started working on chaos theory within the framework of statistical physics. Establishing an intense collaboration with them. With Henk van Beijeren and collaborators he worked on methods inspired by the Boltzmann equation on how to calculate Lyapunov exponents for standard models in statistical physics, such as the hard-sphere gas and the Lorentz gas. The computed results are in good agreement with simulation results by Harald Posch and co-workers. During this period Bob also published an excellent monograph 'Introduction to Chaos in Nonequilibrium Statistical Mechanics', Cambridge University Press, 1999.

Bob's last scientific works were with Ted Kirkpatrick on additional aspects of long-ranged spatial correlations in non-equilibrium fluids, especially emphasizing how different non-equilibrium fluids are from their equilibrium counterparts. Important physical concepts emphasized included how the statistics of non-equilibrium fluctuations were very different than those in equilibrium fluctuations and how in non-equilibrium fluids there was a type of rigidity that leads to super-fast signal propagation. Finally, Bob concluded his professional career with a magnus opus entitled 'Contemporary Kinetic Theory of Matter', written with Henk van Beijeren and Ted Kirkpatrick as co-authors, Cambridge University Press, 2021.

In conclusion we must also remark that Bob had many interests outside of science. He always loved music and through the years he played a variety of musical instruments and was a member of a band. He had a deep interest in seventeenth century Dutch art, and the paintings of Johannes Vermeer in particular. Bob was also very involved in Jewish life and traditions.

First and foremost, Bob was a wonderful person, who had a deep interest in everybody he encountered in his life. All Bob's friends, colleagues, and collaborators will dearly miss him. We were fortunate to have had him as a companion in our lives.

T.R. Kirkpatrick[a,*], J.V. Sengers[a], H. van Beijeren[b]
[a] *University of Maryland, USA*
[b] *Utrecht University, the Netherlands*

[*] Corresponding author.